# REGIONAL END-SYSTOLIC CIRCUMFERENTIAL STRAIN DEMONSTRATES REDUCED FUNCTION IN REMOTE MYOCARDIUM AFTER ANTERIOR STEMI

Steve W. Leung[1], Kanjit Leungsuwan[1], Ahmed Abdel-Latif[2], and Jonathan F. Wenk[1,3]
[1] Gill Heart and Vascular Institute, Division of Cardiovascular Medicine, University of Kentucky, Lexington, KY, USA, {steve.leung@uky.edu, kanjit.leungsuwan@uky.edu}
[2] Department of Medicine, Division of Cardiology, University of Michigan and Ann Arbor VA Healthcare System, Ann Arbor, MI, USA, aalatif@umich.edu
[3] Department of Mechanical and Aerospace Engineering, College of Engineering, University of Kentucky, Lexington, KY, USA, jonathan.wenk@uky.edu

### SUMMARY

Anterior ST-segment elevation myocardial infarction (STEMI) is associated with severe adverse remodeling and increased mortality rates. In this study, anterior STEMI patients were treated with primary percutaneous coronary intervention and then imaged with cardiac magnetic resonance 1-2 days later. The cine images were analyzed with computational feature-tracking algorithms to assess end-systolic circumferential (ESC) strain. It was found that contractile function is not only depressed in the infarct and border zone regions, but also in the remote myocardium. This is contrary to other types of infarctions where remote myocardium shows enhanced compensatory function. Quantifying ESC could help identify high risk patients.

## 1 INTRODUCTION

Anterior ST-segment elevation myocardial infarction (STEMI) exhibits a higher mortality rate and more severe complications compared to other types of STEMI [1]. Thus, there is a pressing need for better prognostic tools and treatment strategies for these patients. Cardiac magnetic resonance (CMR) imaging is widely used for evaluating structural and functional changes in patients post-STEMI [2]. Additionally, computational image analysis techniques (such as feature-tracking algorithms) provide a means for assessing these images to extract information related to myocardial deformation and performance. Several studies have explored the application of left ventricular (LV) strain patterns, measured from CMR images, for evaluating myocardial dysfunction and predicting long-term recovery [3-5]. However, data regarding the regional distribution of strain in patients with anterior STEMI is lacking. Thus, the goal of the present study was to assess end-systolic circumferential (ESC) strain, using computational feature-tracking algorithms to analyze cine images, to evaluate regional function in patients with anterior STEMI compared to normal controls.

## 2 METHODOLOGY

The methods employed in this study are based on those used in our previous study [6]. They are summarized briefly below. This study was conducted using retrospective data from 30 patients with anterior STEMI (N=24 with ejection fraction (EF) < 50% and N=6 with EF ≥ 50%) who were enrolled between January 2014 and September 2019 at the University of Kentucky hospital. Exclusion criteria included: mechanical intubation, hemodynamic instability with cardiogenic shock, administration of dual antiplatelet therapy, or receipt of thrombolytics. All STEMI patients received primary percutaneous coronary intervention (PCI) and then CMR was performed within 1-2 days. The control patients (N=18) were similar to the STEMI patients in terms of sex and age, had no prior coronary artery disease, and were evaluated by CMR for cardiac function. The study protocol was authorized by the University of Kentucky Institutional Review Board and Ethics Committees and complies with the Declaration of Helsinki.

Images were acquired with a 1.5T magnetic resonance scanner (Magnetom Aera, Siemens Medical, Erlangen, Germany), which used a 12-channel spine coil and an 18-channel body coil. Steady-state free precession (SSFP) cine imaging was performed under breath-holding conditions. The long axis images were taken in the 2-, 3-, and 4-chamber views. The short axis images provided full view of the LV from apex to base (representative parameters: matrix: 256 x 256, field of view: 380mm x 380mm, slice thickness: 8mm, interslice gap: 2mm, repetition time (TR): 3.2ms, echo time (TE): 1.2ms, temporal resolution: 50ms, reconstructed into 25 phases, flip angle: 50°).

Level-3 trained CMR readers performed blinded analyses of the imaging data using CMR42 v5.6.2 (Circle Cardiovascular Imaging, Inc., Calgary, Canada). Endocardial and epicardial contours were drawn by the CMR reader on each long axis and short axis cine image at end-diastole and end-systole. The location of the apex and mitral valve plane were established using the long axis images. The insertion points between the LV and RV were identified in the anterior and inferior regions to help determine the 16 AHA segments. The EF was assessed via 3D reconstruction of the end-diastolic and end-systolic volumes. 3D feature-tracking analysis was used to calculate the circumferential strain at 25 phases throughout the cardiac cycle. The ESC strain in each AHA segment was located at the same time point as the end-systolic volume from the EF calculation. A comparison of the temporal and spatial variation in the circumferential strain is presented for a representative control patient and two STEMI patients in Figure 1.

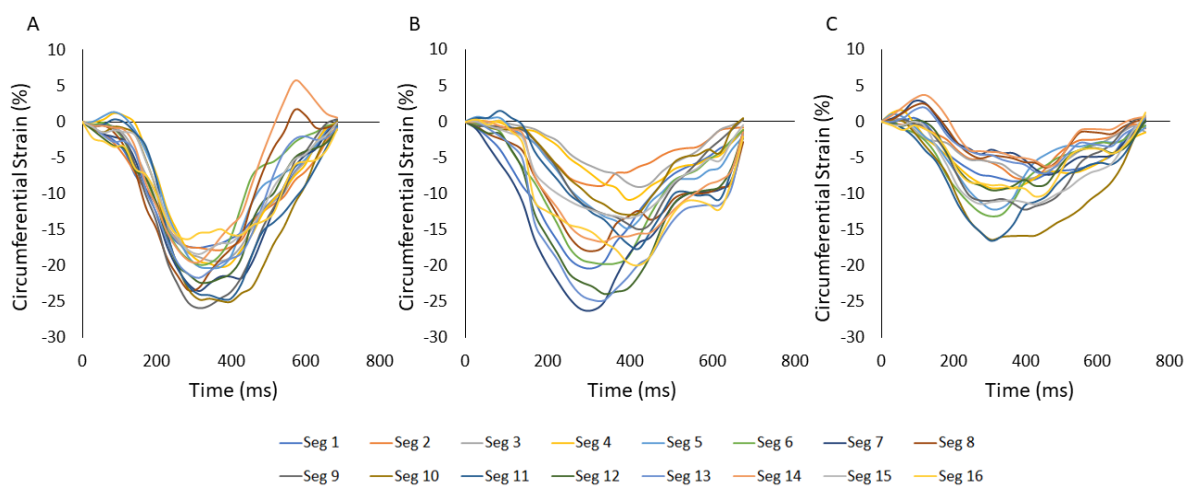

**Figure 1:** Representative circumferential strain vs. time curves measured in (A) control patient, (B) anterior STEMI patient with EF ≥ 50%, and (C) anterior STEMI patient with EF < 50%. Each curve corresponds to the strain in one of the 16 AHA segments of the left ventricle. Note that the peak magnitude of strain decreases as the severity of the STEMI increases from panel to panel.

Depending on the distribution type, data are shown as either mean ± standard error (SE) or median [Q1, Q3]. As noted earlier, the data were separated into 3 groups, which were controls, STEMI patients with EF ≥ 50%, and STEMI patients with EF < 50% (since this typically represents a state of depressed function). Since multiple comparisons are required, analyses were conducted using one-way ANOVA with post-hoc Bonferroni tests. Comparisons between the 3 groups were performed within each of the 16 AHA segments for circumferential strain. The Fisher's exact test was used for comparisons of categorical variables. Significance was identified for a value of $P < 0.05$.

### 3 RESULTS AND CONCLUSIONS
Characteristics for control and STEMI patients are provided in Table 1. Similarities between groups are seen in terms of BMI, age, and sex. The STEMI patients have a higher percentage of diabetes and hypertension compared to the control group. The percentage of smokers was lower in the control group versus the STEMI patients. The median number of days between the initial STEMI and time of CMR was similar between the two STEMI patient groups. All three groups exhibited a significant difference in ejection fraction.

The ESC strain distributions for control and STEMI patients are shown in Figure 2. When comparing STEMI patients with EF ≥ 50% and control patients, it can be seen that 4 segments in the infarct region and 1 segment in the border zone region have magnitudes of strain that are significantly less than the control. However, all the remaining segments, including the remote region, have strain values that are comparable to the control. By contrast, a comparison of STEMI patients with EF < 50% and control patients revealed that 15 of the 16 AHA segments have strain magnitudes that are significantly less than the control. Only a single remote segment had a strain value comparable to the control. Finally, when comparing the two STEMI patient groups, it was found that 11 of the 16 AHA segments had ESC strain magnitudes that were significantly larger in the patients with EF ≥ 50%.

**Table 1:** Patient Characteristics

| Note: (1) Control patients, (2) STEMI patients with EF ≥ 50%, and (3) STEMI patients with EF < 50%. | Control (N=18) | EF ≥ 50% (N=6) | EF < 50% (N=24) | 1 vs. 2 | 1 vs. 3 | 2 vs. 3 |
|---|---|---|---|---|---|---|
| Age (years, ± SE) | 54.3 ± 2.8 | 51.6 ± 3.8 | 53.1 ± 2.2 | 0.64 | 0.74 | 0.77 |
| Male (N, %) | 11 (61%) | 4 (67%) | 21 (88%) | 0.96 | 0.05* | 0.15 |
| Body Mass Index (kg/m$^2$, ± SE) | 28.1 ± 1.5 | 32.7 ± 2.8 | 28.0 ± 1.2 | 0.18 | 0.95 | 0.12 |
| Past Medical History | | | | | | |
|   Hypertension (N, %) | 7 (39%) | 2 (33%) | 17 (68%) | 0.97 | 0.04* | 0.20 |
|   Diabetes (N, %) | 0 (0%) | 4 (67%) | 4 (16%) | <0.01* | 0.07 | <0.01* |
|   Dyslipidemia (N, %) | 2 (11%) | 3 (50%) | 8 (32%) | 0.02* | 0.10 | 0.28 |
|   Coronary Artery Disease (N, %) | 0 (0%) | 1 (17%) | 3 (12%) | 0.06 | 0.13 | 0.67 |
|   Myocardial Infarction (N, %) | 0 (0%) | 0 (0%) | 1 (4%) | 1.00 | 0.39 | 0.65 |
| Tobacco Use | | | | | | |
|   Never Smoker (N, %) | 15 (72%) | 2 (33%) | 8 (33%) | <0.01* | 0.06 | 0.87 |
|   Former Smoker (N, %) | 2 (11%) | 1 (17%) | 2 (8%) | 0.78 | 0.86 | 0.67 |
|   Current Smoker (N, %) | 1 (6%) | 3 (50%) | 14 (58%) | 0.02* | <0.01* | 0.42 |
| Days to CMR (days, IQR [1,3]) | - | 1 [1, 2] | 1 [1, 2] | - | - | 0.50 |
| Left Ventricular EF (%, ± SE) | 64 ± 0.8 | 56.8 ± 2.2 | 35.3 ± 1.4 | <0.01* | <0.01* | <0.01* |

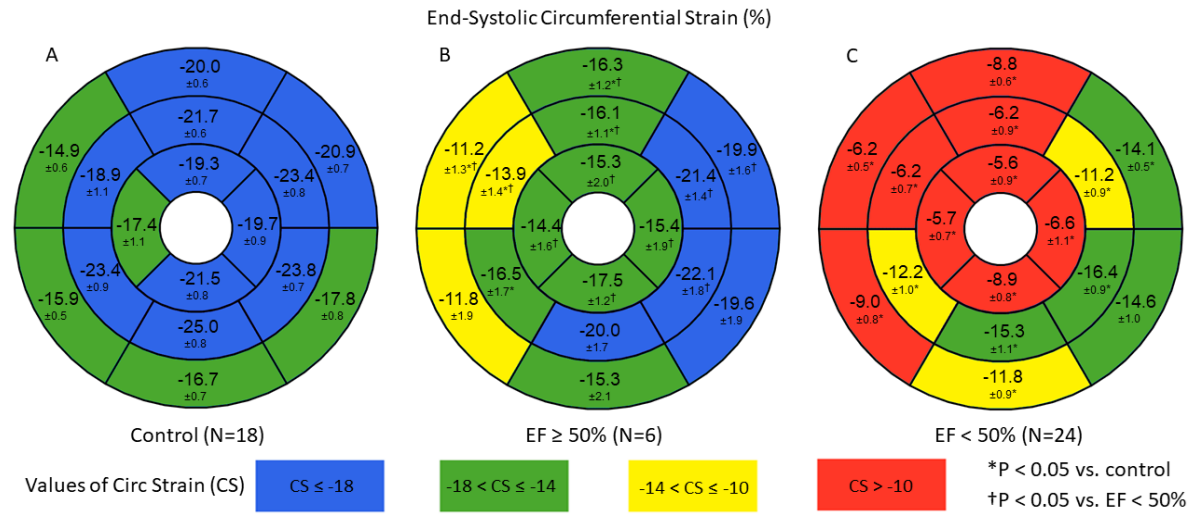

**Figure 2:** End-systolic circumferential strain values in each of the 16 AHA segments for the (A) control patients, (B) anterior STEMI patients with EF ≥ 50%, and (C) anterior STEMI patients with EF < 50%. Comparisons were made within each segment between control patients (N=18), STEMI patients with EF ≥ 50% (N=6), and STEMI patients with EF < 50% (N=24); *P < 0.05 vs. control; †P < 0.05 vs. EF < 50%. The values within each segment are reported as mean ± standard error.

The primary finding from this analysis was that the magnitude of ESC strain was significantly reduced in nearly all segments for anterior STEMI patients with EF < 50%, relative to controls. This implies that the contractile function is not only depressed in the infarct and border zone regions, but also in the remote myocardium of these patients. This agrees with previous studies that used tagged CMR (with coarser spatial resolution) to assess myocardial strain in patients with anterior STEMI [7]. Another observation from the present study was that anterior STEMI patients with EF ≥ 50% exhibited strain values in the remote myocardium that were not elevated relative to controls. Thus, patients with anterior STEMI do not exhibit compensatory contractility in the remote region and, furthermore, show significantly reduced contractility when EF < 50%. This contrasts with our previous study of STEMI in the inferior region [6], where patients with EF ≥ 50% showed elevated magnitudes of strain (compensatory contractility) in the remote region and patients with EF < 50% showed comparable magnitudes of strain in the remote region, compared to controls. This reinforces the notion that anterior STEMI is a much more severe infarct that significantly impairs function over the entire LV. There are several mechanisms that could be contributing to the dysfunction in the remote myocardium of anterior STEMI patients. Microvascular injury has been reported acutely after PCI and could play a significant role in the reduction of circumferential strain seen in the present study [8]. In connection with this phenomenon, the role of inflammation on post-STEMI progression has also been widely studied and linked to adverse remodeling [9]. Additionally, others have studied the effects of geometric alterations on the mechanical loading of the ventricle, which can affect regional wall motion via tethering to the infarct [8].

In conclusion, ESC strain can be used to identify dysfunctional myocardial segments after anterior STEMI. In patients with EF < 50% the severity of dysfunction is more global than in other types of STEMI. Future studies could determine whether regional strain analysis can identify which segments are stunned vs. those with permanent damage. This type of risk stratification could help guide patient-specific treatment strategies.

**4 ACKNOWLEDGEMENTS**
Support for this research was provided by National Institutes of Health grants R01 HL163977 (J.W.) and R01 HL124266 (A.A-L.), as well as the Gill Professorship in Engineering (J.W.).